# Efficient Lightweight Encryption Algorithm for Smart Video Applications

Amna Shifa
*Department of Computer Science & IT,*
*The Islamia University of Bahawalpur,*
Bahawlpur, Pakistan
amnashifa@yahoo.com

Naila Batool
*Department of Computer Science,*
*Govt. Sadiq Women University*
Bahawlpur, Pakistan
anabukhari@live.com

Mamoona Naveed Asghar
*Department of Computer Science & IT,*
*The Islamia University of Bahawalpur,*
Bahawlpur, Pakistan
mamona.asghar@iub.edu.pk

Martin Fluery
*School of CSEE,*
*University of Essex, Colchester, United Kingdom*
fleury.martin55@gmail.com

*Abstract*— The future generation networks: Internet of things (IoT), in combination with advanced computer vision techniques poses new challenges for securing videos for end-users. The visual devices generally have constrained resources in respects to their low computation power, small memory with limited power supply. Therefore, to facilitate videos security in the smart environment, lightweight security schemes are required instead of inefficient existing traditional cryptography algorithms. This research paper provides the solution to overcome such problems. A novel lightweight cipher algorithm is proposed here which targets multimedia in IoT with an in-house name EXPer i.e. Extended permutation with eXclusive OR (XOR). EXPer is a symmetric stream cipher that consists of simple XOR and left shift operations with three keys of 128 bits. The proposed cipher algorithm has been tested on various sample videos. Comparison of the proposed algorithm has been made with the traditional cipher algorithms XOR and Advanced Encryption Standard (AES). Visual results confirm that EXPer provides security level equivalent to the AES algorithm with less computational cost than AES. Therefore, it can easily be perceived that the EXPer is a better replacement of AES for securing real-time video applications in IoT.

## I. INTRODUCTION

A recent expansion of multimedia applications in education, entertainment, medical, surveillance and monitoring, and many more have been growing at a phenomenal rate. Furthermore, the adaptation of multimedia in IoT, particularly live data streaming in home automation, surveillance and mentoring, and intelligent transportation, inherent many security, confidentiality and privacy threats. Therefore, the security of multimedia content i.e. text, images, audio, and animation and video have become a foremost issue. Thus, to ensure the secure end-to-end transmission of information many cryptographic algorithms have been developed up till now. Although, existing ciphers provide higher security, however, they are inappropriate to fulfill the requirements of real-time multimedia applications and fourth generation internet i.e. the internet of things. The conventional cipher algorithms consume higher computation cost in terms of time, memory consumption, bandwidth utilization and processing power. Thus, the limitations of existing ciphers provoke the researchers to develop more efficient lightweight symmetric and asymmetric cipher algorithms. These algorithms are developed with the aim to provide the higher security within the requirements of real-time applications and constraint resources. In this regards, Yao et al. proposed lightweight no-pairing attribute-based encryption by utilizing the Elliptic Curve Decisional Diffie Hellman (ECDDH) for IoT devices [1]. The proposed solution achieved low communication overhead and improved execution efficiency. However, it incurs poor scalability. Then in [2], Xin introduced a hybrid approach by using Elliptic Curve Cryptography (ECC) and the Advanced Encryption Standard (AES) aimed at achieving high-security for an IoT. In the proposed scheme, provides fast and reliable encryption however the complexity and slow execution of ECC adds extra computational overhead. Moreover, lightweight asymmetric such as RSA and Elliptic curve cryptography (ECC) [3] provides higher security but requires more resources, computational time and memory than symmetric algorithms such as AES, HEIGHT, TEA, PRESENT and RC5 [4]. Therefore, symmetric algorithms are preferred in the constrained IoT environment.

Generally, it might be thought that full encryption of compressed video would suffice. However, full encryption for video is often unnecessary, as selective encryption (SE) [5] reduces computation overhead while obscuring the contents of the video. In [6], Kulkarni et al. proposed a new encryption algorithm for videos. In the proposed scheme, firstly by using a video cutter the video is divided into frames. After that frames are shuffled and then passed through a frame stitching block. Then video with shuffled frames is transmitted to the destination along with shuffling key. To attain higher security sensitive codewords are extracted from MVDs, DCs, and ACs and are encrypted using AES. Extraction of selective codewords is done for saving the computation time. However, the AES itself with multiple rounds increases the complexity and execution time. Therefore, this paper proposes an improved lightweight cipher algorithm which can ensure the security of real-time video efficiently. In addition, our proposed scheme contributes to the reduction of the encryption complexity as SE is applied on selected syntax elements that are arithmetic signs of motion vectors, texture coefficients, delta QP and combined motion vectors (signs), texture coefficients (signs) and Delta QP.

The paper is organized as follows: Related work is critically discussed in section II while Section III presents the proposed lightweight cipher algorithm and SE scheme. Experimental results followed by performance evaluation are presented in section IV. The concluding remarks and future work are given in section V.

## II. RELATED WORK

Typical existing methods of cryptography such as AES and DES encryption are inappropriate for real-time applications and videos. Consequently, light-weight and efficient schemes are required to alleviate the large energy consumption, high bandwidth utilization and encryption/decryption overhead. Recently, there has been much interest shown by researchers in designing efficient lightweight algorithms for secure end-to-end communication. In [7], S. Mondal and S. Maitra modified the AES to achieve stronger security and small computational cost. The modification is done by randomizing the key and shifting of pixel position. Shifting of pixels is done row wise and column wise. In typical AES the value of keys is fixed for encryption of an image. So in the modified algorithm, key values are randomized. The analysis shows that the proposed modified form of AES provides better security against statistical attacks. In [8], Charru et al. proposed a cryptography algorithm which is based on XOR operation. In the proposed algorithm, XOR operation is applied to input data. The level of security of an algorithm is represented by computing the number of steps required for decryption. If the number of steps required for decrypting the ciphered text are higher than security will be higher and vice versa. Experimental results show that as compared to the existing algorithm the proposed algorithm takes higher number of steps for decrypting the ciphered text. In [9], the authors discussed an enhancement in the AES algorithm in which dynamic S-boxes were being generated. The modified factor in dynamic AES algorithm is the generation of dynamic S-boxes which are key-dependent, due which the differential and linear cryptanalysis become more difficult. Correlation Coefficient and NIST randomness tests show that dynamic AES was superior to the original one with the verified security. However, dynamic generation of S-boxes and multiple rounds to create the desired random output increase the execution time.

Other recently proposed cipher algorithms for constrained resources include LEA [10], TWINE [11], FeW[12], Prince [13]. However, all of these are based on multiple round structure and their execution efficiency is dependent on the number of rounds. In [14], Z.A. kissel introduced Random Rotating XOR (RRX) cipher algorithm and implementation in both software and hardware effectively. The proposed algorithm is the modified form of XOR. The modification is done by permutation of the key in a random manner. The bits of the key are rotated with a cyclic shift. The rotate function is defined as rot (x,d,b) where x denotes the binary variable, d denotes the direction of rotation and b denotes the number of bits to shift such that $0 < b < n$. After performing bit rotation on key and encryption on the chunk of a message, the encrypted packet is sent to the peer with the information about the direction of rotation and number of bits rotated. Furthermore, to attain higher security the key should not be same for at least two adjacent packets. RRX is a quick and tidy algorithm hence it could be suitable for multimedia security in IoT environment. Recently, in [15], Sundaram et al. used encryption and hash algorithms to achieve security in an IoT. However, the proposed solution fails to deal with large amounts of multimedia big data because of the proposed encryption algorithm's ability to encrypt only 64 bits per block: hence, it suffers from slow performance. Thus, to minimize the encryption time and complexity, in this paper, a lightweight encryption algorithm EXPer is proposed. The proposed algorithm consists of five simple steps/stages with a single iteration over those steps. In each step, simple XOR and shift operations are performed to achieve random output with a low computational overhead.

## III. PROPOSED LIGHTWEIGHT CIPHER ALGORITHM

To ensure the end-to-end secure transmission of video and minimize the encryption overhead a lightweight symmetric encryption algorithm EXPer has been proposed in this research work. EXPER comprises five simple stages with a single iteration over those stages. In each stage, XOR or permutation with shift operation is performed on output generated from the previous stage. EXPer encrypts the bitstream by using XOR with the 128-bit secret key and sub-keys to produce the encrypted data. Moreover, in the proposed algorithm dynamic secret keys (R_key1, R_key2, and R_key3) are generated by using the Pseudo-Random Function (PRF). The permutation is performed by using a right shift operator. Right shift operator shifts the bit pattern of data to the right by the specified offset value. EXPer is considered as stream cipher as it encrypts bitstream by combining the elements of the plaintext bitstream bit-by-bit with secret keys. Fig.1 illustrates the process of proposed lightweight cipher algorithm EXPer.

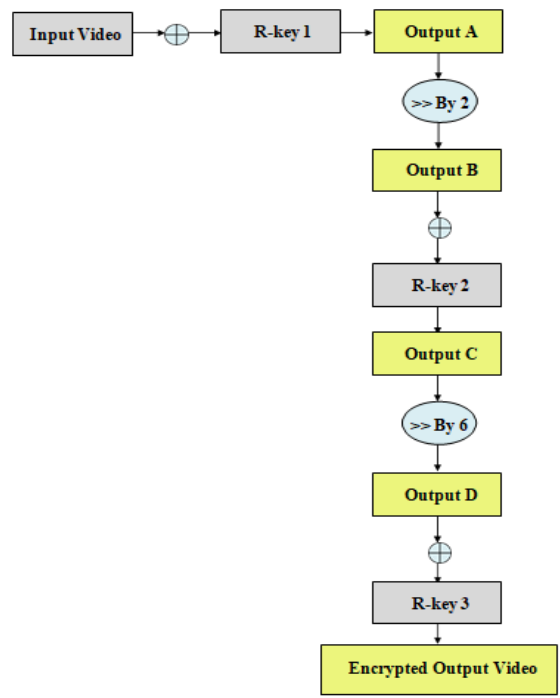

Fig. 1. Working of proposed lightweight Cipher EXPer

The Fig.1 demonstrate that in the first stage, the input bitstream is encrypted by applying XOR operation with round-key1 (R-key1) which is generated by Rand () function, then, the output of first stage is permutated by right shift operator with an offset value of $v_i = 2$ at the second stage. Initially, the offset value is selected 2 to permute the bits within the byte. However, it can be a random number from 1 to 8 to shuffle the bits within the byte only. After that in the third stage, the output of the second stage is XORed with round-key 2(R_key2). At fourth stage bits of resultant out of stage three are shifted by right shift operator with offset value $v_f$ = 6. Finally, at fifth stage, the output of the fourth stage is XORed with round-key 3 (R_key3) which produces the final encrypted output video of higher randomness.

Additionally, to minimize computational complexity, selective encryption (SE) is applied on selected syntax elements that are arithmetic signs of MV, signs of DCT, value of dQP and combined MV (signs), DCT (signs) and dQP. These parameters are selected for the encryption to maintain the overall statistics and compression efficiency after performing SE. Furthermore, these syntax elements are selected to avoid bitrate overhead. The encryption is applied by using in-compression approach on selected elements with proposed algorithm EXPer during Context Adaptive Binary Arithmetic Coder (CABAC) entropy coding within H.264/Advanced Video Coding (AVC) standardized codec, which itself less computationally complex than a High Efficiency Video Coding (HEVC) codec. The Fig.2 illustrates SE based entropy coding with EXPer.

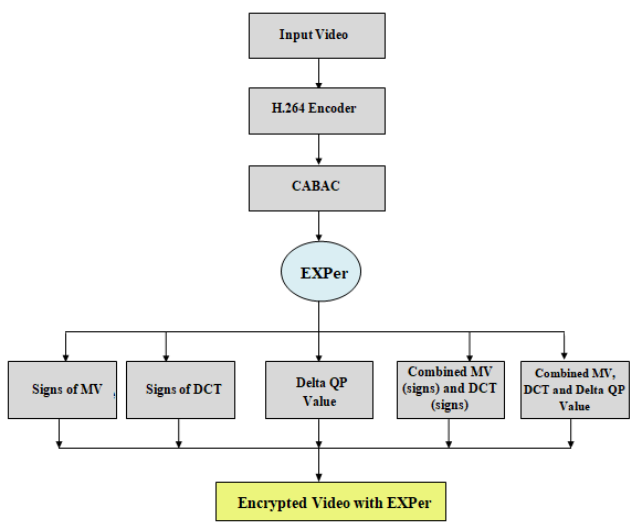

Fig. 2. SE based CABAC entropy coding within H.264/AVC standardized codec with EXPer

As mentioned above EXPer comprises a single iteration with five stages within it, therefore easy to implement. The diffusion or permutation at each stage ensures the high security with low computational complexity. The SE is applied with CABAC because CABAC is more compression efficient and produces less encryption overhead. Fig.3 demonstrates the implementation of SE on selected elements with EXPer during CABAC entropy coding within H.264/AVC codec in the form of pseudo-code.

```
                Pseudo-code of SE with EXPer
1.  Start.
2.  Input : Plain video.
3.  Rand () ; Generate round-keys using Pseudo Random Function (PRF)
4.  Switch (encryption_rank)
5.  {
6.     case Without_Enc: break;
7.        case Medium_Enc:
8.           if (motion_vectors)
9.              {
10.                motion_vector.EXPerEnc_Code();
11.             }
12.          elseif (color_Coeff)
13.             {
14.                coefficient.EXPerEnc_Code();
15.             }
16.           break;
17.       case High_Enc:
18.          if (motion_vectors && color_Coeff)
19.             {
20.                motion_vector.EXPerEnc_Code();
21.                coefficient.EXPerEnc_Code();
22.             }
23.          break;
24. }
25  main()
26. {
27.    int EXPerEnc_Code()
28.       {
29.          uiSign = uiSign ^ R_key1;
30.          uiSign = uiSign >> Vi;
31.          uiSign = uiSign ^ R_key2;
32.          uiSign = uiSign >> Vf;
33.          uiSign = uiSign ^ R_key3;
34.       }
35. }
36. Output: Encrypted Video
37. End
```

Fig. 3. Pseudo-code of SE with EXPer

## IV. EXPERIMENTAL RESULTS AND DISCUSSIONS

In this section performance of the proposed algorithm EXPer is evaluated by applying SE on Common Intermediate Format (CIF) (352×288 pixels/frame) benchmark videos at 30 fps. The SE is applied on signs of MV, signs of DCT, dQP and on combined MV (signs), DCT (signs) and dQP with H.264/AVC at quantization parameter QP= 24. The algorithm is developed using the C/C++ programming language and implemented in the JSVM reference software on a 64-bit operating system with 2.40 GHz Core i5-6200U processor and 8 GB RAM. Sample experimental results of SE with EXPer are shown in Fig.4 and Fig.5. Fig.4 (a) depicts that video is encoded without applying encryption. Fig.4 (b) presents encoded video where only signs of MV are encrypted with EXPer algorithm, Fig.4(c) presents the encoded video where only signs of DCT are encrypted, Fig.4 (d) presents the encoded video where only values of dQP are encrypted, Fig.4 (e) depicts the encoded video where combined MV(signs) and DCT(signs) are encrypted and Fig.4 (f) depicts the encoded video where all three parameters i.e MV (signs) and DCT(signs) and dQP are encrypted. The visual results of Fig.4 (b–f) show that with the proposed algorithm high randomness is achieved without generating considerable encryption overhead. An interesting thing in the experiments to note that the EXPer do not change much in dQP values that's why encryption of dQP is successfully done by this lightweight encryption.

Fig.5 shows the impact results of SE, by using EXPer algorithm at QP= 24, 36 and 48 respectively where all selected elements i.e. signs of MV, signs of DCT and dQP values are encrypted.

In the both Fig. 4 and Fig. 5 the quality distortion within the video sequences can clearly be seen, which imply higher confidentiality is achieved with the proposed cipher EXPer.

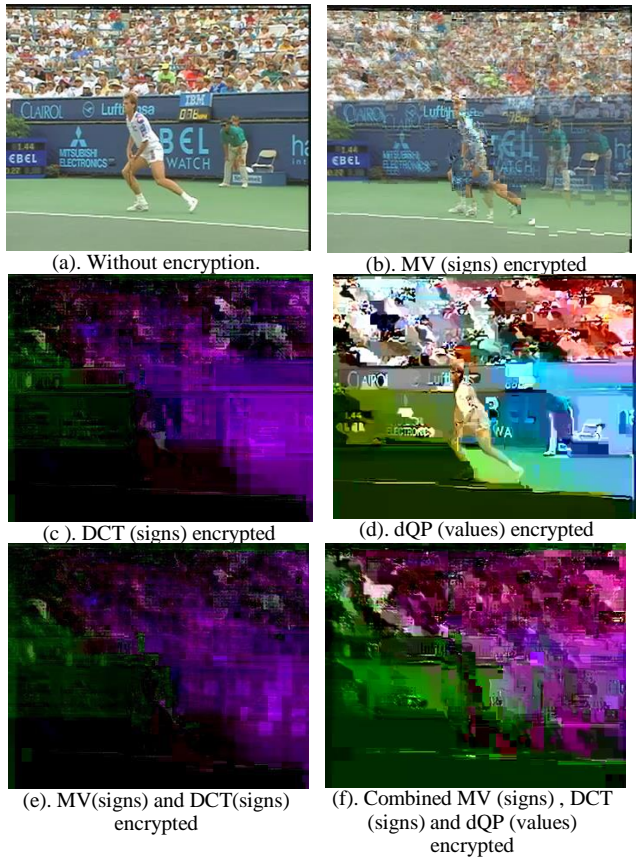

(a). Without encryption.  (b). MV (signs) encrypted

(c). DCT (signs) encrypted  (d). dQP (values) encrypted

(e). MV(signs) and DCT(signs) encrypted  (f). Combined MV (signs), DCT (signs) and dQP (values) encrypted

Fig. 4. Impact of SE (I,B,P Frames) on video parameters with EXPer at QP = 24 on Stephen video.

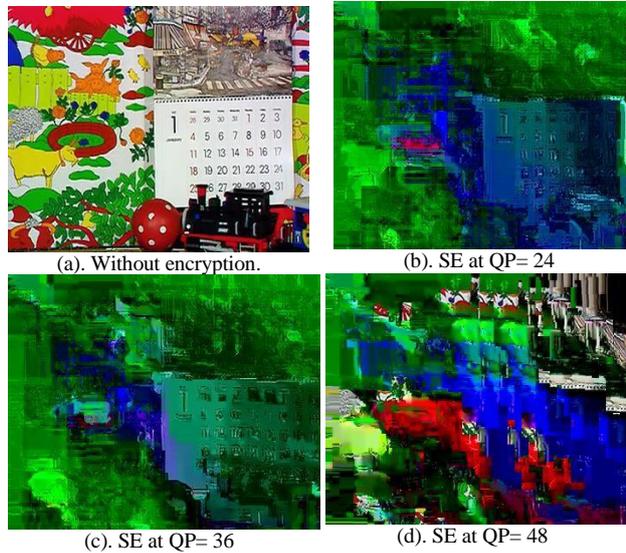

(a). Without encryption.  (b). SE at QP= 24

(c). SE at QP= 36  (d). SE at QP= 48

Fig. 5. Impact of SE (I,B,P Frames) with EXPer at different QP on Mobile video.

Additionally, comparative results are taken by applying SE with XOR and AES algorithms on MV(signs), DCT (signs) and combined MV(signs) and DCT(signs) and dQP at QP=24 and 48 respectively. The comparative visual results are illustrated in Fig.6, where Fig.6 (a) depicts video is encoded(compressed) without applying encryption, Fig.6 (b) represents a video frame that is selectively encrypted with XOR cipher, Fig.5(c) represents a video frame that is selectively encrypted by using AES cipher, and Fig.6 (d) represents video that is selectively encrypted using EXPer cipher. It is noted here that dQP values are not being encrypted with AES cipher during experimentation (Fig. 6 (c)). AES encryption on dQP, has taken dQP values out of the range which crashes the decoder.

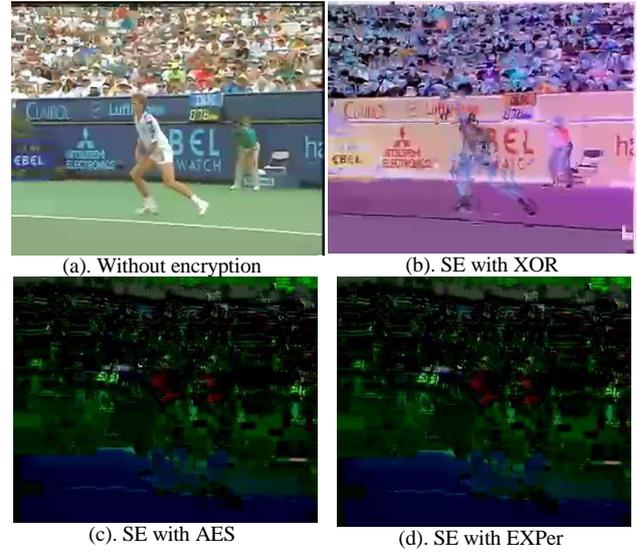

(a). Without encryption  (b). SE with XOR

(c). SE with AES  (d). SE with EXPer

Fig. 6. Comparative impact of SE (I,B,P Frames) on combined MV (signs) and DCT (signs) with EXPer , XOR and AES at QPs = 24 on Stephen video.

From the above visual comparative results, it can be seen that video encrypted with the XOR algorithm is somehow viewable but the video encrypted using EXPer and AES are very difficult to recognize. Therefore, the results imply that the performance of the EXPer algorithm is almost equivalent to the AES algorithm. Moreover, EXPer is so simple that it cannot change much in dQP values that's why encryption of dQP is successfully done by this lightweight encryption which cannot be done by complex state-of-art cipher like AES. Furthermore, Table 1 shows PSNR of test videos encrypted with the EXPer algorithm is very close to the AES algorithm which confirms that EXPer provides security equivalent to AES algorithm. Thus, EXPer algorithm can be considered for the security of real-time video application instead of AES algorithm.

TABLE I. COMPARATIVE PSNR OF SELECTIVE ENCRYPTION WITH EXPER AND AES

| Videos | PSNR (dB) | |
|---|---|---|
| | *AES* | *EXPER* |
| Stephen | [Y=5.76,U=15.6,V=16.97] | [Y=5.96,U=16.69,V=17.04] |
| Football | [Y=10.04,U=18.06,V=19.36] | [Y=0.16,U=19.06,V=20.36] |
| Mobile | [Y=6.07,U=13.18,V=10.79] | [Y=6.19,U=14.18,V= 11.79] |
| Silent | [Y=5.80,U=20.97,V= 26.32] | [Y=6.00,U=21.97,V= 27.32] |

To further evaluate the efficiency of the proposed encryption algorithm, comparative analysis with the encoding time of AES algorithm has been performed. The graph in Fig. 7 clearly indicates that the encoding time of the EXPer is relatively less than AES. Hence, the efficiency of the proposed algorithm (EXPer) in terms of execution time is much better than AES.

As the results confirm that the proposed algorithm provides the same level of security as AES provides but has a negligible computational complexity. Thus, the proposed algorithm could be considered a suitable cipher algorithm for the security of real-time multimedia applications in IoT.

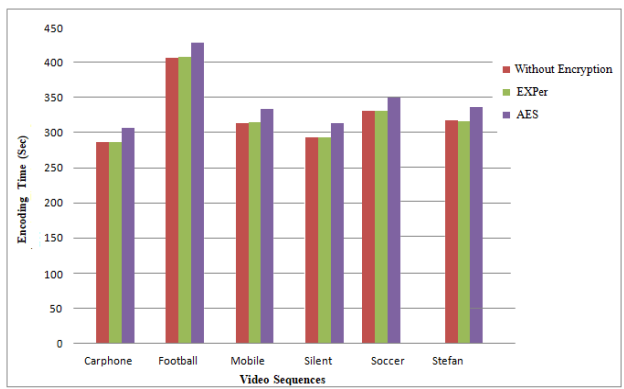

Fig. 7. Compartive Encoding time of EXPer with AES .

## V. CONCLUSION

In this paper, a lightweight cipher algorithm for the end-to-end secure transmission of video in real-time applications has been presented. The proposed encryption algorithm provides better security and efficiency in term of encoding time. The results confirm the visual security level of EXPer is almost compatible with AES, while the computation cost is relatively negligible. Furthermore, the SE adopted in this paper for video security provides a greater level of quality distortion without any encryption bitrate overhead. The proposed solution provides a high level of compression with format compliance, hence a suitable choice for the real-time application for future networks.